\documentclass[aps,prd,twocolumn,a4paper]{revtex4-1}
\usepackage{ulem}
\usepackage{color}
\usepackage{graphicx}
\usepackage{epsfig}
\newcommand{\be}{\begin{equation}}
\newcommand{\ee}{\end{equation}}
\newcommand{\ba}{\begin{eqnarray}}
\newcommand{\ea}{\end{eqnarray}}
\newcommand{\nn}{\nonumber\\}
\begin{document}
\title{Impact of momentum space anisotropy on heavy quark-dynamics in a QGP medium} 
\author{Vinod Chandra $^a$}
\email{vchandra@iitgn.ac.in}
\author{Santosh K. Das $^{b,c}$}
\email{dsantoshphy@gmail.com}
\affiliation{$^a$ Indian Institute of Technology Gandhinagar, VGEC Campus, Ahmedabad-382424, India}
\affiliation{$^b$ Department of Physics and Astronomy, University of Catania, 
Via S. Sofia 64, 1-95125 Catania, Italy}
\affiliation{$^c$ Laboratori Nazionali del Sud, INFN-LNS, Via S. Sofia 62, I-95123 Catania, Italy}

\date{\today}
\begin{abstract}
Momentum space anisotropy present in the quark and gluon distribution functions in
relativistic heavy ion collisions induces Chromo-Weibel instability in the hot QCD medium created therein. 
The impact of the Chromo-Weibel instability on the dynamics of a heavy-quark (HQ) traversing in the QGP medium is
investigated within the framework of kinetic theory by studying the momentum and temperature behavior of HQ drag and diffusion coefficients.
The physics of anisotropy is captured in an effective Vlasov term in the transport equation. The effects of the instability are handled by 
making a relation with the phenomenologically known jet quenching parameter in RHIC and LHC. Interestingly, the  presence of instability 
significantly affect the temperature and momentum dependences of the HQ drag and diffusion coefficients. 
These results may have appreciable impact on the experimental observables such as, the nuclear suppression 
factor, $R_{AA}(p_T)$, and the elliptic flow, $v_2(p_T)$, of heavy mesons in heavy ion collisions at RHIC and LHC
energies which is a matter of future investigation. 

\vspace{2mm}

\vspace{2mm}
\color{black}
\end{abstract}
\maketitle

\section{Introduction}
The physics of the Chromo-Weibel instability~\cite{chromoweibel} 
(non-Abelian analogue of Weibel instability~\cite{Weibel}) during the hydrodynamic expansion of the QGP in 
heavy ion collisions, may play crucial role in understanding the space-time evolution and properties of quark-gluon plasma medium.
The momentum anisotropy present during the hydrodynamic expansion of the QGP may  induce instabilities to the Yang-Mills field (Chromo field) equations.
The Weibel type of  instabilities can be seen in the  expanding quark-gluon
plasmas, since the  width of the momentum component in the direction
of the expansion narrows by expansion, leading to an anisotropic momentum distribution. 
The instability in the rapidly expanding QGP in heavy ion collisions may lead to the 
plasma turbulence~\cite{bmuller}. Recall, the plasma turbulence describes a random, non-thermal pattern of
excitation of coherent color field modes in the QGP with a power spectrum similar to that of 
vortices in a turbulent fluid~\cite{bmuller}.

The prime goal here is to investigate the heavy quark dynamics in the presence of 
Chromo-Weibel instability. This could be done by first modeling  the non-equilibrium momentum distribution 
functions that describe expanding anisotropic QGP followed by employing it to the kinetic theory description
of heavy quark dynamics.

Hadrons containing HQs ($c$, $\bar{c}$, $b$, or $\bar{b}$)
are of great interest in investigating the properties of the QGP, since their physical properties 
get significantly modified while traveling through QGP. This fact has been 
reflected in the particle spectra at RHIC, and the LHC energies.
Further, HQ thermalization time is larger than gluons and light quarks,
and they do not constitute the bulk of the QGP. Since their formation occurs in the 
early stages of the collisions, they can travel through the thermalized QGP medium, and can 
retain the information about the interaction with them very efficiently. For instance,
it is pertinent to ask whether a single $c \bar{c}$ can stay together long enough to 
form a bound state (say $J/\psi$) at the hadronization state. To address this,
one requires to describe the dynamics of the  HQs propagating through the QGP. 
Therefore, one can explore the physics of the HQ transport~\cite{10,mooreteaney,11,12,13,Das0,14,15,16,17,18,19,rappprl,fs,Das,HB,Das1,Younus:2015kba} 
in the QGP medium as follows.The non-equilibrated HQs can travel in the equilibrated QGP medium, and one  has to 
deal the problem within the framework of Langevin dynamics~\cite{fp}. This is to say that the 
HQs perform random motion in the equilibrated QGP. Recall, that 
the QGP goes through a hydrodynamic evolution  before it reaches the hadronization and 
subsequently the hadrons freeze-out.

The pertinent question to ask is, whether a HQs maintain equilibrium during this 
entire process of the space-time evolution or not. It has been observed ~\cite{jane} within the framework of Langevin dynamics 
and pQCD (perturbative QCD) that the HQs may not achieve the equilibrium in the RHIC and LHC energies. 

The most important observable which encode the medium effects carried with them by the HQs
while traveling in the QGP, is the nuclear modification factor, $R_{AA}$. It has been 
observed that their energy loss in the QGP due to gluon radiation is  insufficient to  
describe the medium modification of the spectrum~\cite{bair,jeon}. 
Therefore, one has to look at the collisions since they have different fluctuation 
spectrum than radiation, and might contribute significantly as one thought off 
initially~\cite{coll,akdm}. The collisional 
effects can be captured well in the HQ drag and diffusion coefficients which have been 
calculated within weak coupling QCD by several authors. The formalism, and details are offered 
in Sec. IIA.

The temperature, $T$ and chemical potential, $\mu_B$ dependence of the 
drag and diffusion coefficient enter through the thermal distributions of light quarks and gluons.
In the present calculation, we ignore the $\mu_B$ dependence in view of the fact that the QGP produced at RHIC 
and LHC energies at the mid-rapidity region has negligibly small net baryon density. 
But one has to implement the realistic QGP EoS in terms of appropriate form of the thermal distribution functions. 
Lattice QCD EoS may be a good choice for the description of the  equilibrated QGP. Additionally, it is important to address the role 
of the momentum anisotropies at RHIC and LHC in influencing the dynamics of the heavy quarks in the hot QCD medium.
This is the main focus of the article.

The paper is organized as follows. Section II deals with the kinetic theory formulation of 
HQ dynamics in the background QGP medium in terms of drag and diffusion coefficients. Sec. III, 
discusses the non(near)-equilibrium modeling of the degrees of freedom that describes the QGP medium in the 
presence of anisotropy. In Sec. IV, we present the results and related discussions. Finally, conclusions are 
presented in Sec. V.

\section{Heavy-quark drag and diffusion in the hot QCD medium}
HQs play crucial role in characterizing QGP as they are produced in the early stages 
of the heavy-ion collisions and  remain extant throughout the evolution and hence can capture the 
information of the entire evolution of the system. The dynamics of HQs while traveling in the QGP medium 
can be understood in terms of the drag and diffusion coefficients following Landau's prescription.

\subsection{Heavy Quark drag and diffusion}
Let us  consider the elastic interaction experienced 
by HQs while traversing in to the hot QCD medium. 
Next, we consider the process  $c(p) + l(q) \rightarrow c(p^\prime) + l(q^\prime)$ 
($l$ stands for gluon and light quarks and anti quarks).

\subsection{HQ drag}
The the drag coefficient,  $\gamma$ can be calculated by using the following 
expression~\cite{BS}:
\begin{equation}
\gamma=p_iA_i/p^2
\end{equation}
where $A_i$ is given by 
\begin{eqnarray}
\label{eqn2}
A_i=\frac{1}{2E_p} \int \frac{d^3q}{(2\pi)^3E_q×} \int \frac{d^3p^\prime}{(2\pi)^3E_p^\prime×}
\int \frac{d^3q^\prime}{(2\pi)^3E_q^\prime×}  \nonumber \\ \frac{1}{g_Q} 
\sum  \overline{|M|^2} (2\pi)^4 \delta^4(p+q-p^\prime-q^\prime) \nonumber \\
{f}(q)(1\pm f(q^\prime))[(p-p^\prime)_i] \equiv \langle \langle
(p-p^\prime)\rangle \rangle
\label{eq0}
\end{eqnarray}
$g_Q$ being the statistical degeneracy of the HQ propagating through QGP.
The above expression indicates that the drag coefficient is the
measure of the thermal average of the momentum transfer, $p-p^\prime$ due to
interaction of the heavy quarks with the bath particle weighted by the square of 
the invariant amplitude, $\overline{\mid M\mid^2}$.
The factor $f(q)$ denotes the thermal distribution of the particles
in the QGP. $1 \pm f(p^\prime)$ is  the 
momentum distribution  with Bose enhancement or Pauli suppressed 
probability in the final state. Note that  $f(q)$ will involve three types of thermal phase space distribution functions
corresponding to the  gluons ($g$), light-quarks ($q\equiv$ up and down) and 
the strange quarks ($s$) and corresponding anti-quarks. Hence, $f(q)$ jointly denote these three phase space distribution as,
\begin{equation}
f(q)\equiv \lbrace f_g, f_q, f_s \rbrace .
\end{equation}

In the presence of initial momentum anisotropy, we need to model them 
appropriately by first setting up the transport equation and then solving it 
either analytically or numerically. In the present work, we consider the linearized transport equation 
and capture all the effects coming from the anisotropy as the first order modification to the 
equilibrium distribution functions for quark-antiquark and gluons.

In view of the the above, we consider the following decomposition 
for the f(q) in three sectors\\
\begin{eqnarray}
\label{eqa}
 f_g=f_0^g ( p)+f_1^g(\vec{p}, \vec{r})\nonumber\\
 f_q=f_0^q (p)+f_1^q(\vec{p}, \vec{r})\nonumber\\
 f_s=f_0^s (p)+f_1^s(\vec{p}, \vec{r}).
\end{eqnarray}
Here $p=\vert \vec{p} \vert$.

At this stage, we need the correct modeling of equilibrium (isotropic) distribution 
functions (first term in the right hand side of Eq. (\ref{eqa}) and the modifications induced by the 
anisotropy. This is presented in the next section.

\subsection{HQ diffusion}
Similar to the HQ diffusion coefficient, $B_0$ can be evaluated as:
\begin{equation}
B_0=\frac{1}{4}\left[\langle \langle p\prime^2 \rangle \rangle -
\frac{\langle \langle (p.p\prime)^2 \rangle \rangle }{p^2}\right].
\label{diffusion}
\end{equation}

With an appropriate choice of ${\cal F}(p^\prime)$ both the drag and diffusion coefficients can be
evaluated from a single expression as follows:

\begin{eqnarray}
\label{eqn6}
<<{\cal F}(p)>>=\frac{1}{512\pi^4×} \frac{1}{E_p} \int_{0}^{\infty} \frac{q^2 dq d(cos\chi)}{E_q} \hat{f}(q) \nonumber \\
\frac{w^{1/2}(s,m_Q^2,m_p^2)}{\sqrt{s}} \int_{1}^{-1} d(cos\theta_{c.m.})  \nonumber \\ 
\frac{1}{g_Q} \sum  \overline{|M|^2} \int_{0}^{2\pi} d\phi_{c.m.} {\cal F}(p^\prime)
\label{transport}
\end{eqnarray}
where\, $s$ is the Mandelstam variable
and $w(a,b,c)=a^2+b^2+c^2-2ab-2bc-2ac$ is the triangular function.
In the next section we will modeling of non-equilibrium  distribution functions for a rapidly expanding plasma 
in the presence of small momentum anisotropy.

\section{Modeling momentum distribution functions for gluons and quarks}
\subsection{The isotropic distributions}
The equilibrium modeling of the momentum distribution functions employed here
is based on the quasi-particle nature of the hot QCD medium (beyond $T_c$)~\cite{chandra_quasi}. 
The quasi-particle description employed here, has been developed in the context
of the recent (2+1)-flavor lattice QCD EoS~\cite{cheng} at physical quark masses. There are more recent lattice results with the improved 
actions and refined lattices~\cite{leos1_lat}, for which we need to re-look the model 
with specific set of lattice data specially to define the effective gluonic degrees of freedom.
Therefore, we will stick to the set of lattice data utilized in the model described in~\cite{chandra_quasi}.
Here, form of the equilibrium distribution functions,  $ f_{eq}\equiv \lbrace f_0^{g}, f_0^{q}, f_0^{s} \rbrace$ 
(this notation will be useful later while writing the transport equation in both the sector in compact notations), 
describing the strong interaction effects in terms of effective fugacities $z_{g,q}$ can be written as.

\ba
\label{eq1}
f_0^{g/q} &=& \frac{z_{g/q}\exp[-\beta p]}{\bigg(1\mp z_{g/q}\exp[-\beta p]\bigg)},\nn
f_0^{s} &=& \frac{z_q\exp[-\beta \sqrt{p^2+m_s^2}]}{\bigg(1+z_q\exp[-\beta \sqrt{p^2+m_s^2}]\bigg)},
\ea
where $p=|\vec{p}|$,  $m_s$ denotes the mass of the strange quark(which we choose to be $0.1 GeV$),
and $\beta=T^{-1}$ denotes inverse of the 
temperature. 

We use the notation $\nu_g=2(N_c^2-1)$ for gluonic degrees of freedom,
$\nu_{q}=2\times 2\times N_c\times 2$ for light quarks, $\nu_s=2\times 2 \times N_c \times 1$ 
for the strange quark for $SU(N_c)$. As we are working 
at zero baryon chemical potential, therefore quark and antiquark distribution functions are the same.=
Since the model is valid in the deconfined phase of QCD (beyond $T_c$), therefore, the mass of the light quarks can be neglected as compared to 
the temperature.  As QCD is a $SU(3)$ gauge theory so $N_c=3$ for  our analysis. 

Note that the effective fugacities ($z_{g/q}$) are not merely a temperature
dependent parameter which encodes the hot QCD medium effects. They lead to 
non-trivial dispersion relation both in the gluonic and quark sectors as,
\ba
\label{eq2}
\omega_g&=&p+T^2\partial_T ln(z_g)\nn
\omega_q&=&p+T^2\partial_T ln(z_q)\nn
\omega_s&=&\sqrt{p^2+m^2}+T^2\partial_T ln(z_q),
\ea
and this lead to the new energy dispersions for gluons ($\omega_g$), 
light-quark antiquarks ($\omega_q$),
and strange quark-antiquarks. A detailed discussions of the interpretation and physical significance of $z_g$, and $z_q$, we refer the reader 
to~\cite{chandra_quasi}. There are other quasi-particle descriptions in the literature, those could be 
characterized as, effective mass models ~\cite{effmass1,effmass2},
effective mass models with gluon condensate~\cite{effmass_glu}, and 
effective models with Polyakov loop~\cite{effmass_pol}. 
Our model is fundamentally distinct from all these models.
Another crucial point is regarding the definition of the 
energy momentum tensor, $T^{\mu\nu}$. As described in~\cite{dusling}, in the presence of non-trivial temperature 
dependent energy dispersion (as in all these quasi-particle models), we need to modify the definition of the 
$T^{\mu\nu}$ so that the trace anomaly effects in QCD can be accommodated in the definition. The 
modified $T^{\mu\nu}$ for the effective mass models is obtained in~\cite{dusling}, and for the current model 
in ~\cite{chandra_new}.

\subsection{Chromo-Weibel instability and anomalous transport: Dupree-Vlasov equation}
 Recall, the momentum anisotropy present in quark and gluon momentum distribution functions 
 induces instability in the Yang-Mills equations in similar way as Weibel instability in the case of 
 Electromagnetic  plasmas. This instability while coupled with the rapid expansion of the QGP leads to anomalous transport and 
 modulates the transport coefficients of the plasma substantially. This fact is realized by Dupree in the case of EM plasmas in 1954~\cite{dupree} and
 later generalized for the non-Abelian plasmas in ~\cite{bmuller,chandra_eta1}. In the context of QGP, the phenomenon of the anomalous transport 
 is realized at the later stages of the collisions as due to the hydrodynamic expansion of the QGP, one  has appreciable momentum anisotropy present in 
 thermal distribution functions of quark and gluons. 
 
 The first step towards estimating the near equilibrium distributions function for the 
 quarks and gluons in rapidly expanding QGP with momentum anisotropy, is to 
 set up the Dupree-Vlasov equation (linearized version) and then solve with the help of an ansatz to obtain the correction to the 
 isotropic distribution functions. Here, we briefly outline the mathematical formalism in solving the transport equation.
 
 \subsection{Formalism}
 We start with the following ansatz for the non-equilibrium distribution function,
 \begin{equation}
 \label{eqf}
  f(\vec{p}, \vec{r})=\frac{z_{g,q} \exp(-\beta u^\mu p_\mu)}{1\pm z_{g,q} \exp\big(-\beta u^\mu p_\mu +f_1 (\vec{p}, \vec{r})\big)}
 \end{equation}
 $z_{g,q}$ are the effective gluon, quark fugacities coming from the isotropic modeling of the QGP in terms of lattice QCD equation of state. The parameter 
 $\beta$ is the temperature inverse (in units of $K_B=1$), $u^\mu$ is the fluid 4-velocity considering fluid picture of the QGP medium. 
 Here, $f_1 (\vec{p}, \vec{r})$ denotes the effects from the anisotropy (momentum). To achieve the above mentioned near equilibrium situation, 
 $f_1$ must be a small perturbation. Under this condition, we obtain,
 \begin{equation}
  f(\vec{p}, \vec{r})=f_0(p)+f_0 (1\pm f_0 (p)) f_1(\vec{p},\vec{r}) +O(f_1(\vec{p}, \vec{r})^2).
 \end{equation}
The {\it plus} sign is for gluons and {\it minus} sign is for the quarks/antiquarks.

Next, the following form for the ansatz is considered for the linear order  perturbation to the isotropic  gluon and quarks distribution functions respectively,
\begin{equation}
f_1 (\vec{p},\vec{r})\equiv f_1^{g,q}= -\frac{1}{\omega_{g,q} T^2} p_i p_j (\Delta_{g,q} (\vec{p}) (\nabla u)_{ij}, 
\end{equation}
The quantities, $\Delta_{g,q}$ denotes the strength of the momentum anisotropy for the gluons and quarks respectively.
In the local rest frame of the fluid (LRF) $f_0=f_{eq}=(f_0^g, f_0^q)$, and considering longitudinal boost invariance, we obtain,
$\nabla\cdot \vec{u}=\frac{1}{\tau}$ and ${\nabla u}_{ij}=\frac{1}{3 \tau} diag (-1,-1,2)$, leading to \\
\begin{equation}
 f_1^{g,q}=-\frac{\Delta_{g,q} (p)}{\omega_{g,q} T^2 \tau} (p_z^2-\frac{p^2}{3}).
\end{equation}
Let us now proceed to set up the effective transport equation in the presence of turbulent Chromo-fields that are induced by the 
momentum anisotropy in the thermal distribution of the quasi-gluons and quarks while coupled with the rapid expansion of the QGP medium. 

\subsubsection{Effective transport equation in turbulent chromo fields}
The evolution of the quasi-quark and quasi-gluon momentum distribution functions 
in the anisotropic QGP medium can be described by the Vlasov-Boltzmann equation~\cite{bmuller}.
After invoking the argument that the soft color fields are turbulent and that their action on the quasi-partons in 
can be described by taking an ensemble average, the  Vlasov-Boltzmann equation
can be replaced by Dupree's ensemble averaged, diffusive Vlasov-Boltzmann 
equation~\cite{bmuller}:
\begin{equation}
v^\mu\frac{\partial}{\partial x^\mu} \bar{f} 
-{\mathcal F_A} \bar{f}= 0 \, .
\label{eq:DVBE}
\end{equation}
Here, $\bar{f}$ denotes the ensemble averaged thermal distribution function of quasi-partons. In our case, 
$\bar{f}\equiv f(\vec{p}, \vec{r})$ (given in Eq. (\ref{eqf})). Note that we are only considering the anomalous transport 
, the collision term is not taken in to account here.

The force term (${\mathcal F_A}$) in the case of Chromo-electromagnetic plasma in the present case will be,
\begin{eqnarray}
 {\mathcal F_A} \bar{f}(p) &&\equiv {\mathcal F_A} f(\vec{p}, \vec{r})\nonumber\\
 &=&   \frac{g^2 C_2}{3 (N_c^2-1) \omega^2_{g,q}} <E^2+B^2> \tau_m\nonumber\\ &&\times {\mathcal L}^2
f_{eq}(1\pm f_{eq}) p_i p_j (\nabla u)_{ij}.
 \end{eqnarray}
 
Where $C_2$ is the Casimir invariants ($C_2\equiv \big(N_c, (N_c^2-1)/{2 N_c}\big)$ quadratic Casimirs of $SU(N_c)$). The quanties $<E^2>$ and $<B^2>$
are the color averaged Chromo-electric and Chromo-magnetic fields (average over the ensemble of turbulent color fields~\cite{bmuller}), 
$\tau_m$ is the time scale  (relaxation time) for the instability. Note that while obtaining effective Vlasov-Dupree equation in~Eq.(\ref{eqt}).
The operator ${\mathcal L}^2$ is defined as:
\begin{equation}
{\mathcal L}^2= [\vec{p}\times \partial_{\vec{p}} ]^2-[\vec{p}\times \partial_{\vec{p}} ]^2_{z}.
\end{equation}
While obtaining the expression for the above force term, we first considered a purely Chromo-magnetic
plasma and then written down the terms in light cone frame~\cite{bmuller,chandra_eta}.

Now, we start with the the equilibrium distribution function (local) 
$f_{eq}=1/(z^{-1}_{g,q}\exp(\beta u.p)\mp 1)$, where $z_{g/q}$ is purely temperature dependent.
The action of the drift operator on $f_{eq}$ is given by
\begin{eqnarray}
\label{eqd}
(v\cdot\partial)f_{eq}&=&-f_{eq}(1+f_{eq})\bigg\lbrace(p-\partial_{\beta} \ln(z_{g,q}))v\cdot \partial (\beta)
\nonumber\\&&+\beta (v\cdot \partial)(u\cdot p)\bigg\rbrace,
\end{eqnarray}
 where we recognize that $p-\partial_{\beta} \ln(z_{g/q})\equiv \omega_{g,q}$,  is the modified dispersion relations.
For us the third term in the right hand side of Eq.(\ref{eqd}) is useful, as we are mainly concerned about the anisotropic 
expansion (other two terms contribute to the thermal conductivity and bulk viscosities respectively).
 
 The final expression for the drift term after imposing the energy-momentum conservation is obtained as
\begin{eqnarray}
\label{eqd} 
(v\cdot\partial)f_{eq}(p)&=&f_{eq}(1\pm f_{eq})\bigg[\frac{p_i p_j}{\omega_{g,q} T} (\nabla u)_{ij}\nonumber\\&&-\frac{m^2_D <E^2> \tau^{el}
\omega_{g,q}}{3T^2 {\partial {\mathcal E}}/{\partial T}}\nonumber\\&&+(\frac{p^2}{3\omega_{g,q}^2}-c^2_s)\frac{\omega_{g,q}}{T}(\nabla\cdot\vec{u})\bigg],\nonumber\\ 
\end{eqnarray}
where $c^2_s$ is the speed of sound, $m^2_D$ is the Debye mass, ${\mathcal E}$ is the energy density,  $\tau_{el}$ is the time scale
of the instability in Chromo-elctric fields.

Finally the effective Vlasov-Dupree equation (linearized) by considering the ensemble of 
turbulent color fields with the above ansatz is formulated in~\cite{bmuller, chandra_eta} reads: 
\begin{eqnarray}
\label{eqt}
 &&\bigg\lbrace (\frac{p^2}{3 \omega_{g,q}}-c_s^2)\frac{\omega_{g,q}}{T} (\nabla\cdot \vec{u})+\frac{p_i p_j (\nabla)_{ij}}{\omega_{g,q} T}\bigg\rbrace f_0^{g,q} (1\pm f_0^{g,q})=\nonumber\\
 &&\frac{g^2 C_2}{3 (N_c^2-1) \omega^2_{g,q}} <E^2+B^2> \tau_m {\mathcal L}^2 f_1^{g,q} (\vec{p}) f_0^{g,q} (1\pm f_0^{g,q})\nonumber\\
 \end{eqnarray}

The operator ${\mathcal L}^2$ is similar to the quadrapole operator and the most peculiar thing about it is that it only picks up the anisotropic piece of 
any function of momentum ($\vec{p}$). Importantly, the first term in the left hand side of Eq.(\ref{eqt}) contribute to the physics of
isotropic expansion (bulk viscosity effects) which is not taken in to account in the present work.

Solving  Eq.~(\ref{eqt}) for $\Delta_{g,q}$ analytically, we obtain the following expression~\cite{chandra_new,chandra_bulk},
\begin{equation}
\label{eqdel}
\Delta_{g,q}=2 (N_c^2-1)\frac{\omega_{g,q} T}{3 C_{g,q} g^2 <E^2+B^2>_{g,q} \tau_m}.
\end{equation}

Next, we relate the unknown quantities in the denominator with the phenomenologically known parameter the jet quenching parameter in both gluonic and quark sector
below.

\subsubsection{Relation to the jet quenching parameter, $\hat{q}$}
The two most relevant transport coefficients related to anomalous transport 
due to the soft color fields are the $\eta$ and the jet quenching parameter $\hat{q}$. Here the strength of the
anisotropy, $\Delta(\vec{p})$ is related to the physics of $\eta$. The $\hat{q}$ is proportional to the mean momentum square per unit length 
on the an energetic parton imparted by turbulent fields~\cite{ask:2010}. This fact has been employed to relate the two below.

In the QGP phase, $\hat{q}$ for both gluons ($\hat{q}_g$) and quarks ($\hat{q}_q$) has been estimated employing several different approaches~\cite{burke}.
The five distinct approaches mentioned in~\cite{burke} are {\it viz.}, GLV-CUJET Model~\cite{55}, Higher Twist Berkeley Wuhan Model (HT-BW)~\cite{50}, 
The Higher-Twist-Majumder Model (HT-M)~\cite{51}, MARTINI Model~\cite{56} and The MCGILL AMY Model~\cite{42}. Combining all these models, one obtains the 
quark transport parameter $\hat{q}_q$ in the range, 

\begin{eqnarray}
\label{eqht}
\frac{\hat{q}_q}{T^3}&=&4.6\pm 1.2 \ \ \textit{at RHIC}\nonumber\\
\frac{\hat{q}_q}{T^3}&=&3.7\pm 1.4 \ \ \textit{at LHC}
\end{eqnarray}

The gluon quenching parameter $\hat{q}_g$ is related to $\hat{q}_g$ by a factor of $\frac{9}{4}$ (in terms of Casimir invariants of the $SU(3)$ group),
\begin{equation}
 \hat{q}_g=\frac{9}{4} \hat{q}_q.
\end{equation}
Relevant point to be noted is that $\hat{q}$ for the QGP scales with $T^3$. If one considers the highest temperatures reached at central Au-Au at RHIC and
Pb-Pb at LHC, $T=370 Mev$ and $T=470 Mev$ respectively. The corresponding numbers for $\hat{q}_q$ for a $10 Gev$ quark Jet are, 
\begin{equation}
 \hat{q}_q=1.3\pm 0.3 \ GeV^2/{fm}; \ 1.9\pm 0.7\ Gev^2/{fm},
\end{equation}
for RHIC and LHC respectively.

Let us now discuss the temperature variations at RHIC and LHC while obtaining $\hat{q}$ enlisted in Eq. (\ref{eqht}). For Au-Au at $200$ GeV/n, 
$T_0=346-373$ Mev and for Pb-Pb at $2.76$ TeV/n, $T_0=447-486$ MeV with initial time $\tau_0=0.6$ fm/c for RHIC energy and $\tau_0=0.3$ fm/c for the LHC energy. 
In the present context, the unknown quantities $<E^2+B^2> \tau_m$ which captures the physics of anisotropy and chromo-Weibel instability~\cite{chromoweibel} can be 
written in terms of $\hat{q}$ both in gluonic and matter sectors as\cite{majumder}
\begin{equation}
\label{eqhat}
 \hat{q}= \frac{2 g^2 C_{g/f}}{2(N_c^2-1)} <E^2+B^2> \tau_m,
\end{equation}
where $C_g= N_c$, $C_f=\frac{(N_c^2-1)}{2 N_c}$ for the gluons and quarks respectively.

Invoking the definition of $\hat{q}$ from Eq. (\ref{eqhat}) in Eq. (\ref{eqdel}, we obtain the following expressions,
\begin{eqnarray}
 \Delta_{g,q}= \frac{4 \omega_{g,q}^2 T}{9 \hat{q}_{g,q}}.
\end{eqnarray}

Finally, we obtain the following near equilibrium distribution functions in terms of the jet quenching parameter $\hat{q}$,
\begin{eqnarray}
\label{eq20}
 f^{g,q} (\vec{p})=f_0^{g,q}-f_0^{g,q} (1\pm f_0^{g,q}) \frac{4 \omega_g}{9 \hat{q}_{g,q} (\tau T)} \big(p_z^2-\frac{p^2}{3}\big).\nonumber\\
\end{eqnarray}

\section{Results and discussions}
\begin{figure}
\begin{center}
\includegraphics[width=17pc,clip=true]{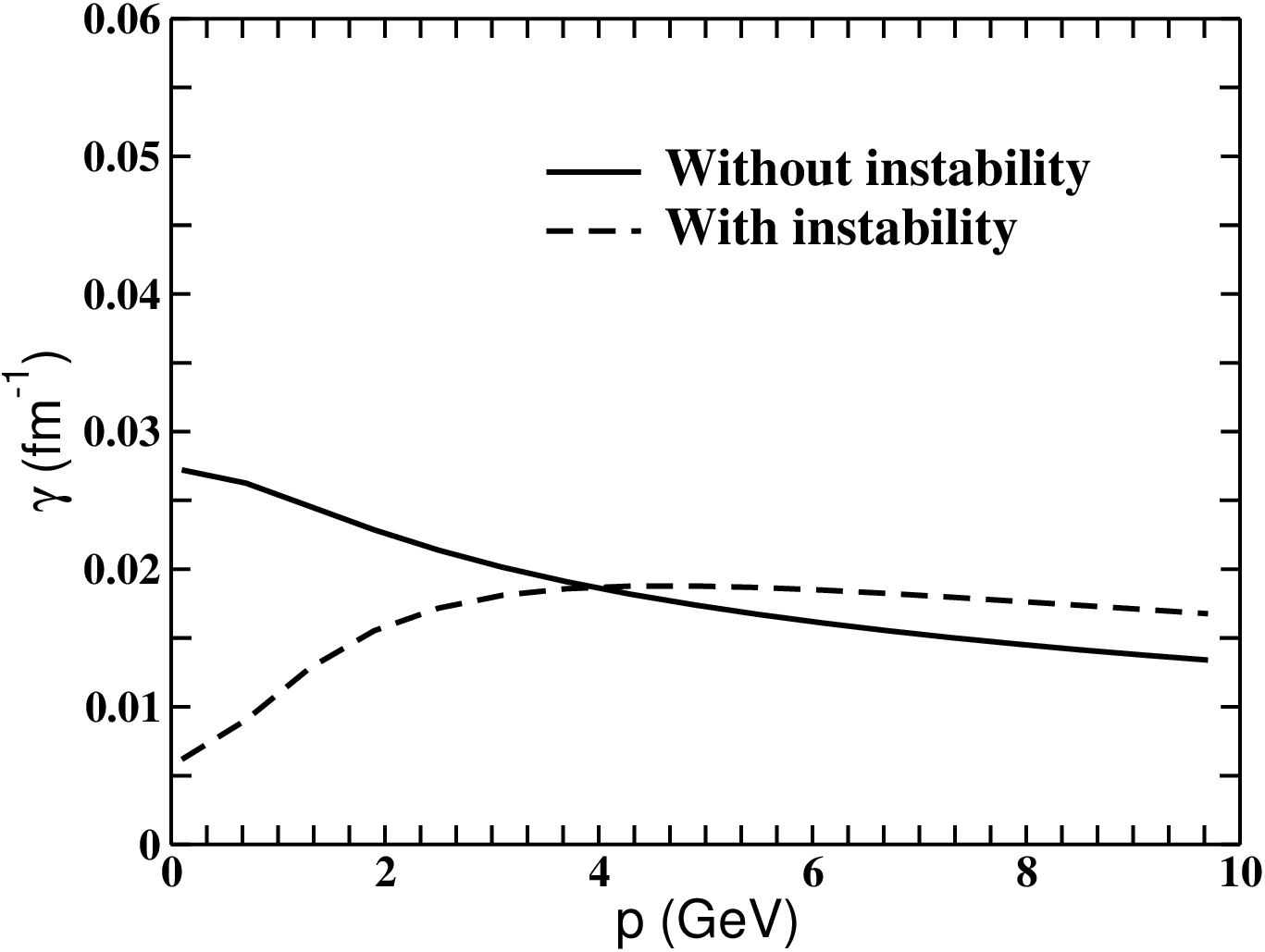}\hspace{2pc}
\caption{Variation of the drag coefficient with momentum at RHIC energy.}
\label{fig1}
\end{center}
\end{figure}

\begin{figure}
\begin{center}
\includegraphics[width=17pc,clip=true]{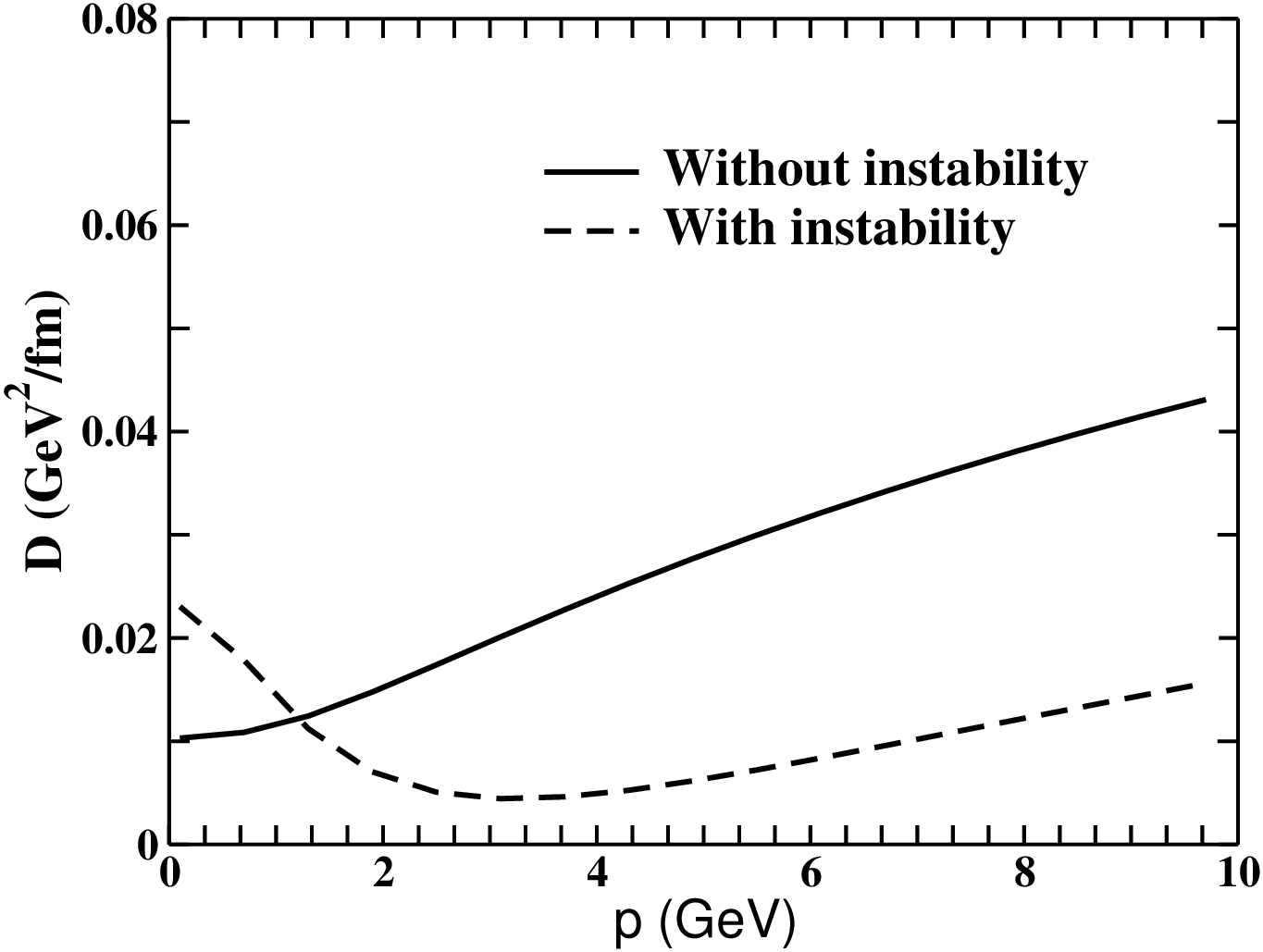}\hspace{2pc}
\caption{Variation of the diffusion coefficient with momentum at RHIC energy.}
\label{fig2}
\end{center}
\end{figure}

The momentum variation of the drag and diffusion coefficients  
of charm quark  has been depicted in~Fig.\ref{fig1} and Fig.~\ref{fig2} with and without instability at RHIC energy by invoking 
Eq. (\ref{eq20}) in Eq. (\ref{eqn2}) and Eq.(\ref{eqn6}) respectively.

The initial temperature ($T_i$) at RHIC energy  assumed to be equal to $T_i=360$ MeV and the $\hat{q}$ corresponding to the 
temperature $360$ MeV is taken as $4.6$. The initial thermalization time ($\tau_t$) at RHIC energy is taken as $0.6$ fm.  
The impact of instability is quite significance (mainly at low momentum range)  which   
decrease the drag coefficient at low momentum, hence, allowing the heavy quarks to move freely . 
It is worth to mention that the temperature dependence of the drag 
coefficient play a significance role~\cite{Das} to describe heavy quark  $R_{AA}$ and $v_2$ simultaneously, 
which is currently a challenge to almost all the models on HQ dynamics. A constant or weak temperature dependence of the 
drag coefficient is essential to reproduce the heavy quarks $R_{AA}$ and $v_2$ simultaneously.
In the presence of instability the drag coefficient decreases 
at high temperature (at low momentum) and it does not affect the low temperature part of the drag coefficient.
Hence, presence of instability alter the temperature as well as the momentum dependence of the drag coefficient 
and may have a significance role on $R_{AA}$ and $v_2$ relation.
We will address these aspects in future works. The variation of the corresponding diffusion coefficient 
with momentum has been shown in ~Fig.\ref{fig2} at the RHIC energy with and without instability. In case of 
diffusion coefficient the impact of instability is noticeable throughout the momentum range 
considered in this work.

\begin{figure}
\begin{center}
\includegraphics[width=17pc,clip=true]{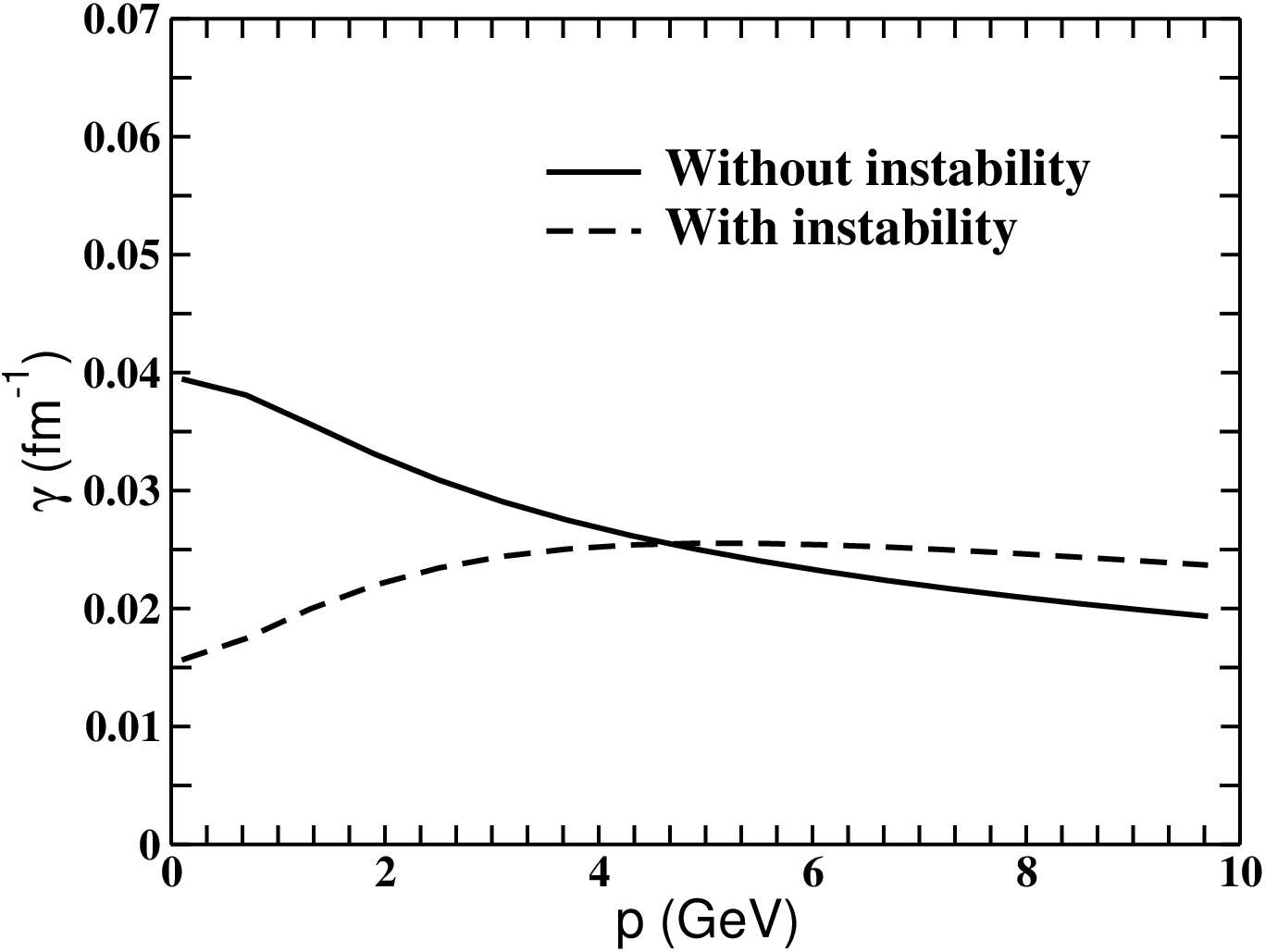}\hspace{2pc}
\caption{Variation of the drag coefficient with momentum at LHC energy.}
\label{fig3}
\end{center}
\end{figure}

\begin{figure}
\begin{center}
\includegraphics[width=17pc,clip=true]{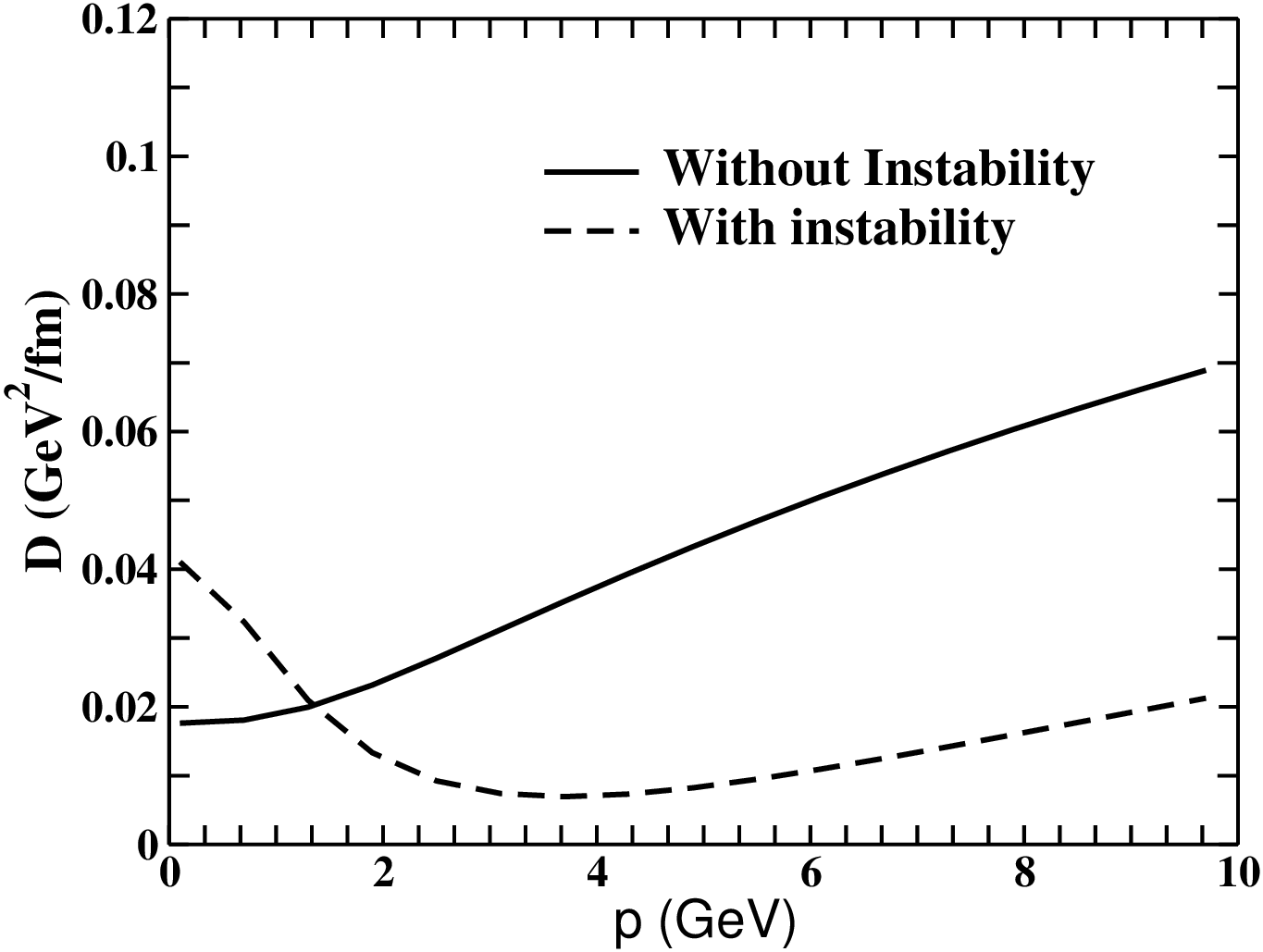}\hspace{2pc}
\caption{Variation of the diffusion coefficient with momentum at LHC energy.}
\label{fig4}
\end{center}
\end{figure}

The momentum variation of drag and diffusion coefficients of charm quarks with and without instability at the
LHC energy are displayed in ~Fig.\ref{fig3} and ~Fig.\ref{fig4} , respectively,
showing behavior qualitatively similar to that of the RHIC energy. In case of LHC energy 
we use $T=480$ MeV and $\hat{q}=3.7$. The initial thermalization time at LHC energy assumed to be $\tau_i=0.3$ fm.
At the qualitative front HQ drag and diffusion coefficients both at RHIC and LHC show similar trend at lower as well as higher 
momentums. This may be due to that fact that the temperature dependence of $\hat{q}$ at RHIC and LHC is not very different.

\subsection{Impact of strength of the anisotropy} 
\begin{figure}
\begin{center}
\includegraphics[width=17pc,clip=true]{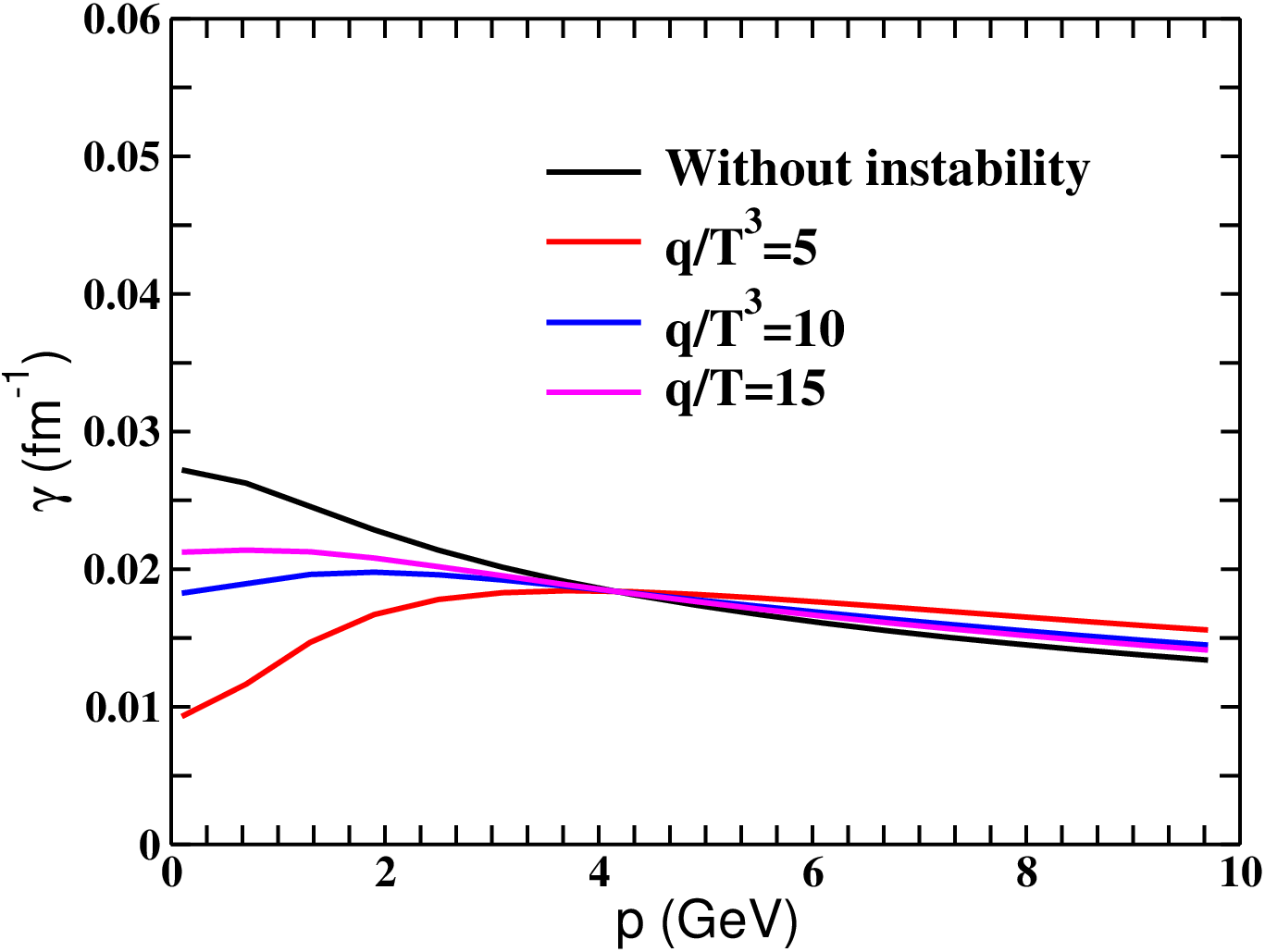}
\caption{Dependence on the strength of the anisotropy/instability of the Drag coefficient at RHIC}
\label{fig5}
\end{center}
\end{figure}

To explore the impact of the instability/anisotropy on 
the heavy-quark dynamics, we vary the 
parameter $\hat{q}/T^3$ from $5-15$ as shown in Fig. \ref{fig5}. As we increase the 
value of $\hat{q}/T^3$, conversely decreasing the  strength 
of the anisotropy, the heavy quark drag coefficient, $\gamma$ at low $p$ (less than 4 GeV) increases, in 
contrast as its behavior at high $p$ (larger than 
4 GeV). The impact is more pronounced at low momentum. Larger the strength of anisotropy, smaller 
the $\gamma$, meaning that the anisotropy is creating relatively lesser hindrance 
for the HQs to travel in the QGP medium, at low momentum,  in contrast, to the role played by the anisotropy at high $p$.

\begin{figure}
\begin{center}
\includegraphics[width=17pc,clip=true]{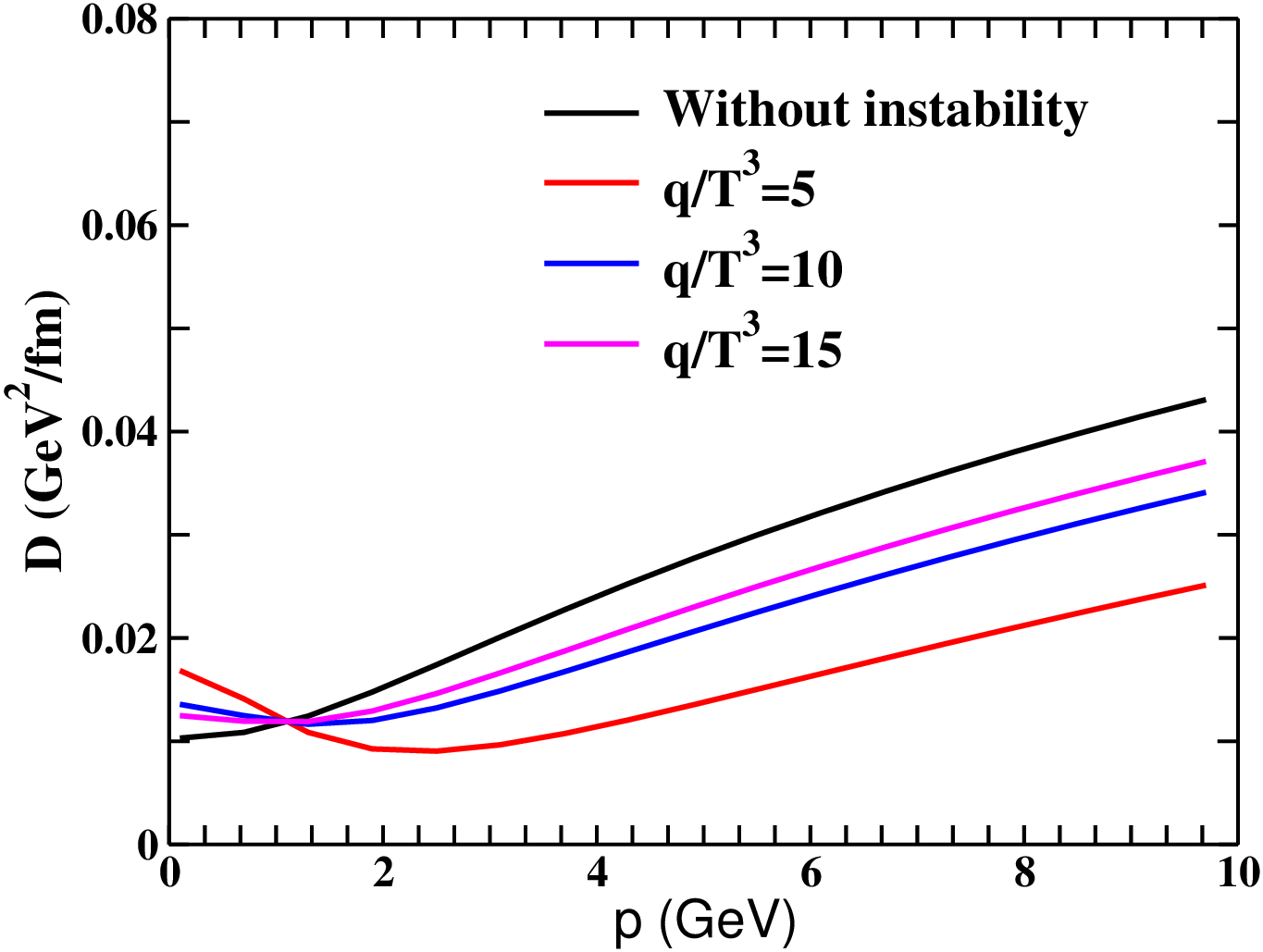}
\caption{Dependence on the strength of the anisotropy/instability of the Diffusion coefficient at RHIC}
\label{fig6}
\end{center}
\end{figure}

We vary the parameter $\hat{q}/T^3$ from $5-15$ as shown in Fig. \ref{fig6} for the diffusion coefficient.
As we increase the value of $\hat{q}/T^3$ the heavy quark diffusion coefficient decreases (in low momentum) 
in contrast to the drag coefficient. The quantity $q$ in the figure legends (Figs. 5-8) is defined as: $q\equiv \hat{q}/T^3$.

\begin{figure}
\begin{center}
\includegraphics[width=17pc,clip=true]{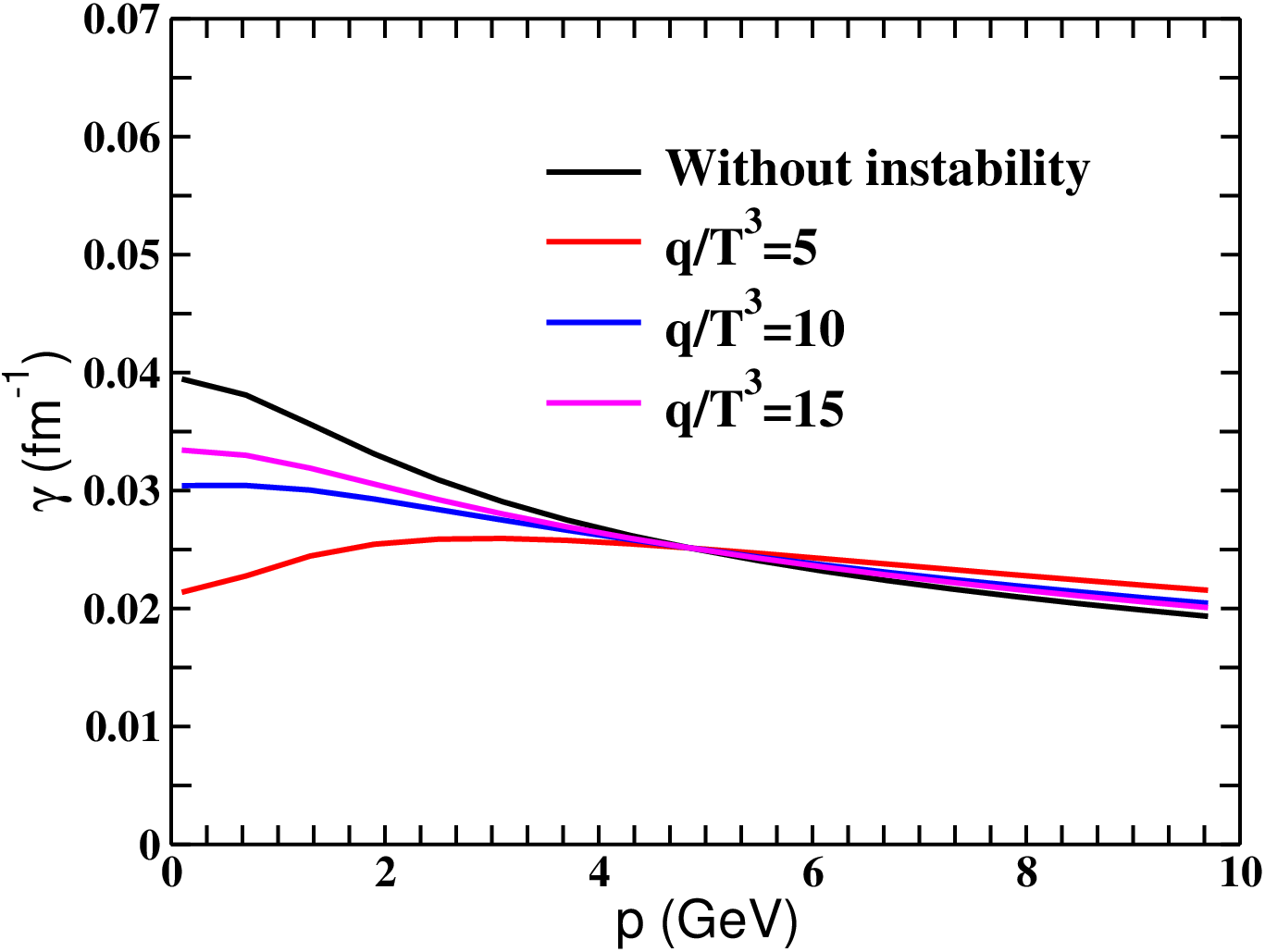}
\caption{Dependence on the strength of the anisotropy/instability of the Drag coefficient at LHC}
\label{fig7}
\end{center}
\end{figure}

\begin{figure}
\begin{center}
\includegraphics[width=17pc,clip=true]{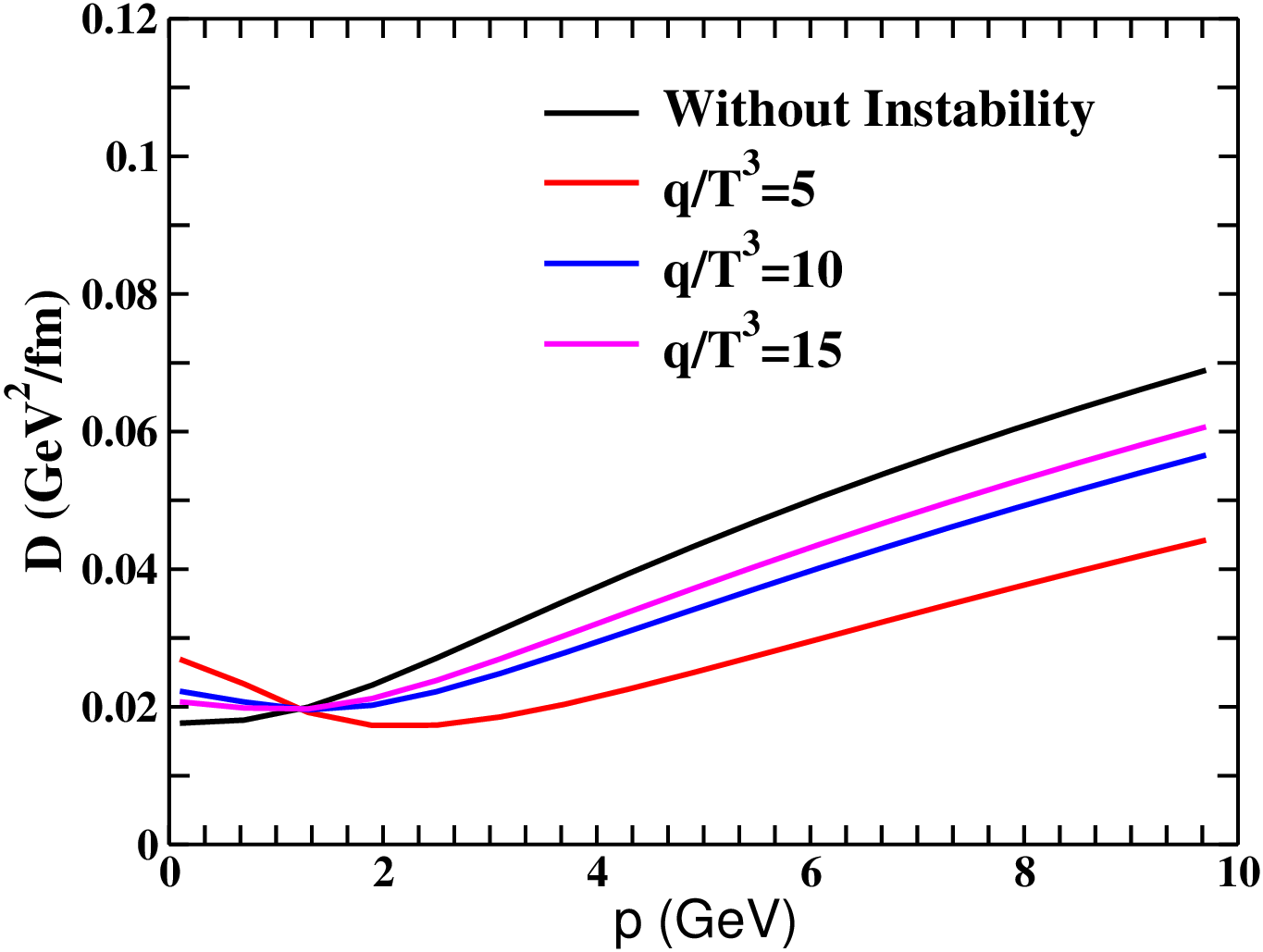}
\caption{Dependence on the strength of the anisotropy/instability of the Diffusion coefficient at LHC}
\label{fig8}
\end{center}
\end{figure}

\section{Conclusions and outlook}
We have estimated the drag and diffusion coefficients of heavy quarks propagating through 
a QGP medium considering the role of momentum state anisotropy. The initial momentum anisotropy 
in the early stages coupled with the rapidly expanding QGP is modeled by setting up  an effective 
transport equation  and its solution in near equilibrium approximation 
leads to the modeling of near(non) equilibrium distribution functions for quark-antiquark and gluons.
We have coupled these distribution functions to the kinetic theory description of heavy quark drag and diffusion 
coefficients and studied their temperature and momentum dependence.

We found that both at RHIC  and LHC energies,  impact of 
the anisotropy on heavy quark transport is quite significant as compared to 
case while HQs are moving in an isotropic QGP medium. The presence of anisotropy alter both the temperature as well as momentum dependences
of the heavy quarks drag and diffusion coefficients. These results  may have  significance impact on $R_{AA}$ and $v_2$ which will be a matter of 
future investigation. We also intend to explore the impact of bulk viscosity along the similar lines of the analysis.

\section*{Acknowledgements} 
This work has been conducted under the INSPIRE Faculty 
grant of Dr. Vinod Chandra ({\tt Grant no.:IFA-13/PH-55, Department of Science and Technology, Govt. of India}) at Indian Institute of Technology Gandhinagar, India. 
We would like to record our sincere gratitude to the people of India for their generous support for the research in basic sciences in the country.
SKD acknowledges the support by the ERC StG
under the QGPDyn Grant no. 259684.

\end{document}